\documentclass[prl, amsmath,amssymb,floatfix,longbibliography,twocolumn,reprint, aps, nofootinbib,]{revtex4-2}%
\usepackage{graphicx}  
\usepackage{dcolumn}  
\usepackage{bm}          
\usepackage[usenames,dvipsnames]{color}
\usepackage{ulem}       
\begin{document}
\title{Coexistence of active and hydrodynamic turbulence in two dimensional active nematics}
\author{C. Rorai$^{1}$, F. Toschi$^{2}$ and I. Pagonabarraga$^{1,3,4}$}
\affiliation{
$^1$CECAM, Centre Europ\'een de Calcul Atomique et Mol\'eculaire, \'Ecole Polytechnique F\'ed\'erale de Lausanne (EPFL), Batochime, Avenue Forel 2, 1015 Lausanne, Switzerland; $^2$Department of Applied Physics, Fluid Dynamics and Heat Transfer, Technical University of Eindhoven (TU/e); $^3$Departament de F\'{\i}sica de la Mat\`eria Condensada, Universitat de Barcelona, C. Mart\'{\i} i Franqu\`es 1, 08028 Barcelona, Spain; $^4$University of Barcelona Institute of Complex Systems (UBICS), Universitat de Barcelona, 08028 Barcelona, Spain.}
\begin{abstract}
In active nematic liquid crystals activity is able to drive chaotic spatiotemporal flows referred to as active turbulence. Active turbulence has been characterized through theoretical and experimental work as a low Reynolds number phenomenon. We show that, in two-dimensions, the active forcing alone is able to trigger hydrodynamic turbulence leading to the coexistence of active and inertial turbulence. This type of flows develops for sufficiently active and extensile  flow-aligning nematics. We observe that the combined effect of an extensile nematic and large values of the flow-aligning parameter leads to a broadening of the elastic energy spectrum that promotes a growth of kinetic energy able to trigger an inverse energy cascade. 
\end{abstract}
\maketitle
 \renewcommand{\baselinestretch}{0.75}
Active fluids, such as microbial suspensions, cytoskeletal suspensions, self-propelled colloids, and cell tissues are complex fluids distinguished by the presence of an active phase whose single units self-propel \cite{Alert21, Marchetti13, Ramaswamy10}. The collective dynamics of the self-driven elements can generate spontaneous flows characterized by chaotic spatiotemporal patterns, referred to as active or mesoscale turbulence \cite{Alert21,  Bratanov15, Dunkel13, Wensink2012}. Active turbulence is subject to intense investigation using a wide variety of theoretical models \cite{Alert21} ranging from phenomenological generalizations of the Navier-Stokes equations (GNS) \cite{Dunkel13} to equations for active liquid crystals \cite{Thampi2016} with either nematic or polar order.
Understanding active turbulence is relevant in biological and ecological systems to describe nutrient mixing and molecular transport at the microscale \cite{Kurtuldu11, Siddhartha21, Ran21}. Identifying the universal features of active turbulence and its similarities with hydrodynamic turbulence where the motion is dominated by the nonlinear terms of the Navier-Stokes equations, namely the inertial terms, remains an open challenge \cite{Alert21}. 

Experimentally, active turbulence is  a low Reynolds number, Re, phenomenon~\cite{Dombrowski04, Wensink2012}. Theoretically, it has been described emerging   from an ensemble of vortices whose area is exponentially distributed and whose isotropic kinetic energy spectrum in two-dimensions follows a $k^{-1}$ or $k^{-4}$ power-law, depending on whether the wavenumber $k$ is smaller or larger than a characteristic vortex size~\cite{Giomi15}.  Initial comparisons with computer simulations  provided convincing evidence of the $k^{-4}$ scaling, while the $k^{-1}$ power-law was later confirmed by high-resolution Stokes flow numerical simulations~\cite{Joanny20}.

Active fluids have also attracted interest for their rheological behavior. Theoretical work showed that, in the regime of linear rheology, activity either thins or thickens the flow depending on the swimmers' propulsion mechanism (pushers vs pullers) and their response to shear (flow-tumbling vs flow-aligning) \cite{Hatwalne04, Giomi10}.  
Notably, experiments~\cite{Lopez15} confirmed the theoretical prediction of an inviscid superfluidlike regime for pusher-type bacterial suspensions, for which active turbulence may be influenced by inertia, a circumstance of interest for $2D$ flows, where the inverse energy cascade can drive flows up to the system-size scale. This consideration opens up new scenarios that challenge  active turbulence as a merely low-Re number phenomenon.

Although a $2D$ modified version of the GNS equation~\cite{Slomka15} has  shown that active turbulence can spontaneously trigger hydrodynamic turbulence~\cite{Linkmann2019}, the active nematohydrodynamic (AN) model~\cite{DeGennes, Thampi2016},  derived from conservation laws and system symmetries, has parameters with a direct physical meaning, allowing  for an insightful understanding of the  mechanisms  causing turbulence.

In this work we report the coexistence of active and hydrodynamic turbulence in $2D$ active nematics. The coupled active-hydrodynamic turbulent state is linked to a  distribution of elastic energy across scales that emerges as the combined effect of an extensile nematics and large  flow-aligning parameters. This feature is intrinsic to active nematics,  even in low-Re number flows; however, it is at intermediate Re numbers that it triggers the inverse energy cascade leading to large Re number flows.

The AN model~\cite{DeGennes, Thampi2016, Rorai21} couples  the evolution of the nematic phase with the incompressible Navier-Stokes equation endowed by a modified stress term. The nematic phase is represented by a second-order tensor: $Q_{ij} = q_0 (n_i n_j - \delta_{ij}/2)$, where $\delta_{ij}$ is the Kroneker delta, $q_0$ is the magnitude of the orientational order, and ${n_i}$ is the director field. The model adopts the Landau-de Gennes free energy 
$\mathcal{F} = \int d^3r [\frac{K}{2}(\partial_k Q_{ij})^2+\frac{A}{2}Q_{ij}Q_{ji}+ \frac{B}{3}Q_{ij}Q_{jk}Q_{ki}+\frac{C}{4}(Q_{ij}Q_{ji})^2],$
where $K$ is the elastic constant.
The relaxation of the orientational order is controlled by the molecular field tensor
$
\mathcal{H}_{ij}  = -\frac{\delta \mathcal{F}}{\delta Q_{ij}}+\frac{\delta_{ij}}{2}Tr\frac{\delta \mathcal{F}}{\delta Q_{kl}}.
$
The equations for the nematic order parameter, $Q_{ij}$, and the velocity field, $u_i$, read
\begin{align}
(\partial_t+u_k\partial_k)Q_{ij}-S_{ij}& =\Gamma \mathcal{H}_{ij},\label{QijEq}\\
(\partial_t+u_k\partial_k)u_i & =\frac{1}{\rho}\partial_j{\Pi}_{ij},\label{NSEq}
\end{align}
complemented by  incompressibility, $\partial_k u_k  = 0$. $\Gamma$ is the rotational diffusivity, $S_{ij}$ the co-rotation term, $\rho$ the fluid density, and $\Pi_{ij}$  the pressure term. The co-rotation term is given by
$
S_{ij}=(\xi E_{ik}+\Omega_{ik})(Q_{kj}+\delta_{kj}/2)\label{Sij}
+(Q_{ik}+\delta_{ik}/2)(\xi E_{kj}-\Omega_{kj})-2\xi(Q_{ij}+\delta_{ij}/2)(Q_{kl}\partial_ku_l),
$
 where $E_{ik}$ and $\Omega_{ik}$ are the strain rate and vorticity tensor, respectively, while  $\xi$ is the flow-aligning parameter and controls whether the active material is flow-aligning or flow-tumbling. The pressure term is given by the sum of an hydrodynamical, passive nematic and active nematic contribution:
$\Pi_{ij}=\Pi_{ij}^{hydro}+\Pi_{ij}^{passive}+\Pi_{ij}^{active}$, where
$\Pi_{ij}^{hydro} = -P\delta_{ij}+2\eta E_{ij}$,  $\Pi_{ij}^{passive} = 2\xi(Q_{ij}+\delta_{ij}/2)(Q_{kl}\mathcal{H}_{lk})-\xi \mathcal{H}_{ik}(Q_{kj}+\delta_{kj}/2)
-\xi(Q_{ik}+\delta_{ik}/2)\mathcal{H}_{kj}-\partial_iQ_{kl}(\delta \mathcal{F}/\delta \partial_jQ_{lk}) +Q_{ik}\mathcal{H}_{kj}-\mathcal{H}_{ik}Q_{kj}$
and $\Pi_{ij}^{active} = -\alpha Q_{ij}$, with $\alpha$ the activity parameter; $\alpha \geq 0$ ($\leq 0$) for a contractile (extensile) active nematic.

\begin{table}[!h]
\caption{Simulation parameters in a system of size $L$. The grid and time spacing are taken as reference units. $T_{fin}$ is the final simulation  time, $Re^*$ is the nominal Re number, $Re^* = \sqrt{K\alpha}q_0 L/(\nu^2\rho)$, and  $l_a=\sqrt{K/|\alpha|}$ is the nominal active length, see SM. 118 simulations were run.}

\begin{center}
\tabcolsep=0.11cm
\begin{tabular}{|c|c|c|c|c|c|}
\hline
Group&$L$ & $T_{fin}$ & $Re^*$ & $l_a$ & $\xi$ \\
\hline
A & 560 & $10^{7}$ & 0.055 & -3.26, -2.45, -1/2,  & 0, 1, 2 \\
  &         &  &                         & 1/2, 0.87, 1,  & \\
    &         &  &                         & 1.73, 2.45, 3.26 & \\
\hline
B &560 & $5\cdot10^{5}$ & 5.5 & 0.5, 0.87, 1.0,  & 0, 1, 2 \\
 & & &  &1.73, 2.45, 3.26 &  \\
\hline
B &560 & $5\cdot10^{5}$ & 5.5 & -6.5, -6.0, -5.5 &  1 \\
\hline
B &560 & $5\cdot10^{5}$ & 5.5 & -1/2, -1, -2.45,  &  2 \\
 & &  &                                      &-3.26, -3.5, -3.75 &  \\
\hline
B &560 & $5\cdot10^{5}$ & 5.5 & -1/5, -2.45, -3.26,  &  0,1 \\
 &&  &  &-4.5, -5 &   \\
\hline
C &560 & $5\cdot10^{5}$ & 22 & -3.26, -2.45, -1,  &  0,1,2 \\
 & &  &  &-1/2, 1, 2.45, 3.26&   \\
\hline
C &560 & $5\cdot10^{5}$ & 22 & -1/5 &  0,1 \\
\hline
C &560 & $5\cdot10^{5}$ & 22 & -0.4, -0.3, -1/4 &  2 \\
\hline
C &560 & $5\cdot10^{5}$ & 22 & 1/2 &  0 \\
\hline
D & 2240 & $10^{7}$ & 0.22 & 0.87, 2.45 & 0 \\
\hline
D & 2240 & $\ge3 \times 10^{7}$ & 0.22 & -1/2, 1/2 & 0, 1/2, 1, \\
 &  &  &  &  &  3/2, 2 \\
\hline
E & 2240 & $10^{6}$ & 22 & 1/2, 0.87, 1,  & 0 \\
 & &  &  & 1.73, 2.45, 3.26 &  \\
\hline
E & 2240 & $2\cdot10^{6}$ & 22 & 1/2, 2.45 & 1 \\
\hline
E & 2240 & $2\cdot10^{6}$ & 22 & 1/2, 0.87, 1, & 2 \\
&  & &  & 1.73, 2.45 &  \\
\hline
F & 6144 & $1.03\cdot10^{6}$ & 60 & -1/2, 1/2 & 2 \\
\hline
\end{tabular}
\end{center}
\label{Runs}
\end{table}%
\begin{figure*} \centering
\begin{minipage}{1.0\textwidth}
\centering
\resizebox{11cm}{!}{\includegraphics{Fig1a}}
\hspace{0.5cm}
\resizebox{5.5cm}{!}{\includegraphics{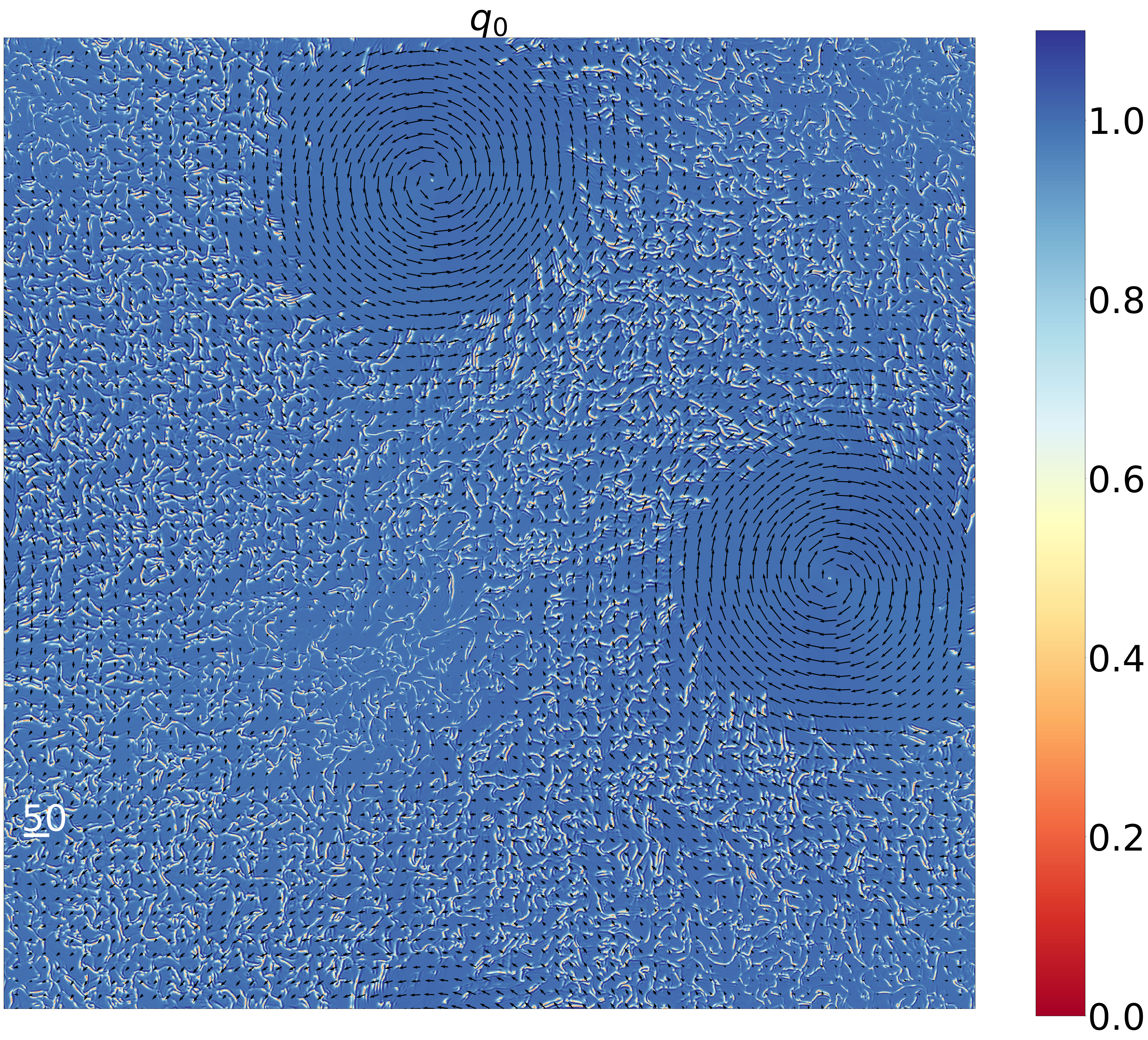}}
\put(-505,125){(a)}
\put(-350,125){(b)}
\put(-170,125){(c)}
\end{minipage}%

\caption{\label{fields}  Snapshots of the magnitude of the velocity field rescaled by the root-mean-square velocity over the entire system (same scale) for the contractile (a) and extensile (b) nematic of group F at time $T_{fin}$, see Table \ref{Runs}. See the SM for visualizations of the vorticity and scalar order parameter. (c) Magnitude of the tensor order parameter (in color) and velocity field (black arrows) for the last time step of the run of group E with $\alpha>0$, $l_a=0.5$ and  $\xi=2$. Unlike contractile nematics (a), the velocity field for extensile nematics (b)-(c) generates large scale structures that  span all the system size. See movies and SM. The white segment in the three panels is of 50 unit length.} 
\end{figure*}
\begin{figure*} \centering
\resizebox{8cm}{!}{\includegraphics{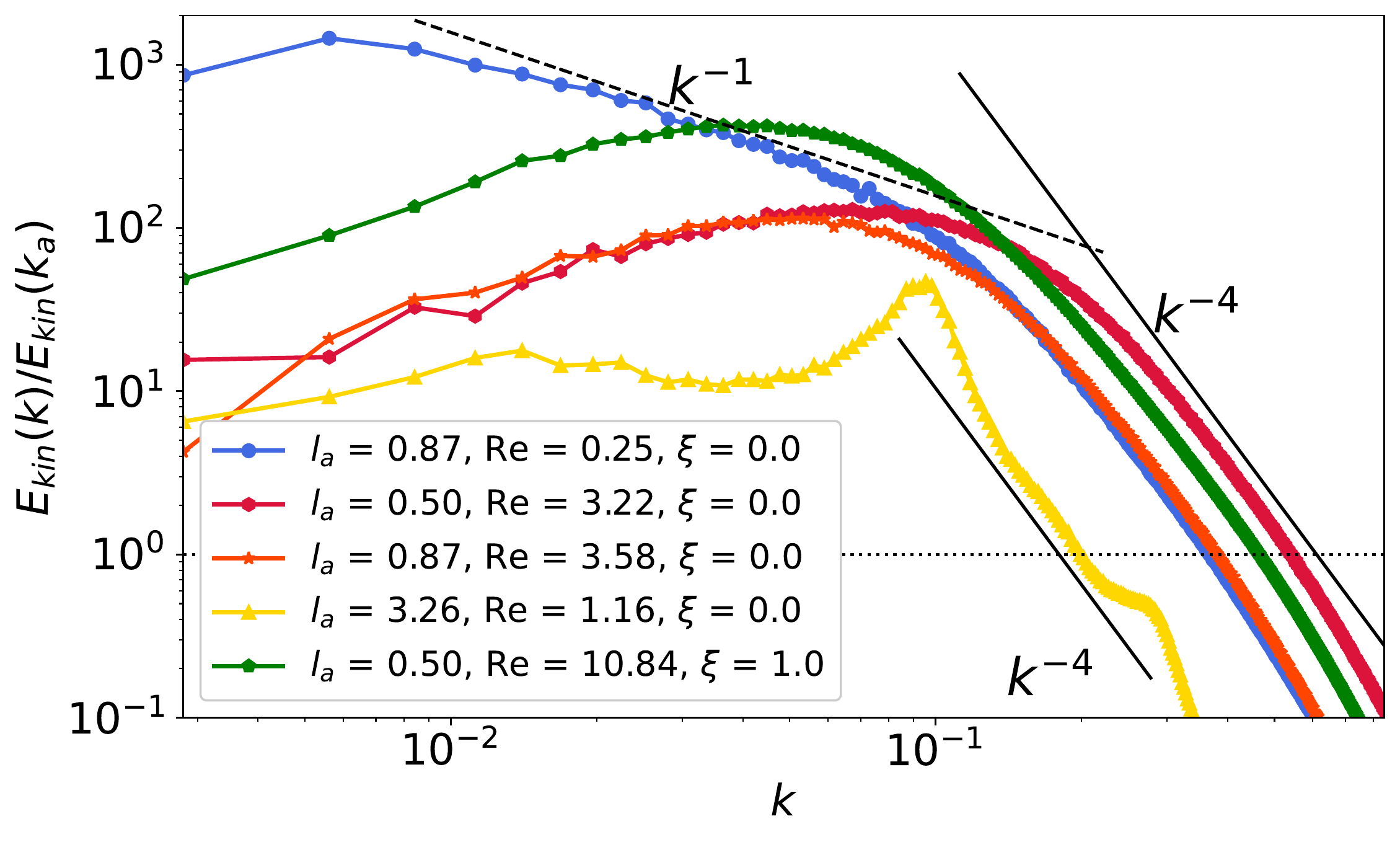}}
\put(-230,135){(a)}
\hspace{0.33cm}
\resizebox{8cm}{!}{\includegraphics{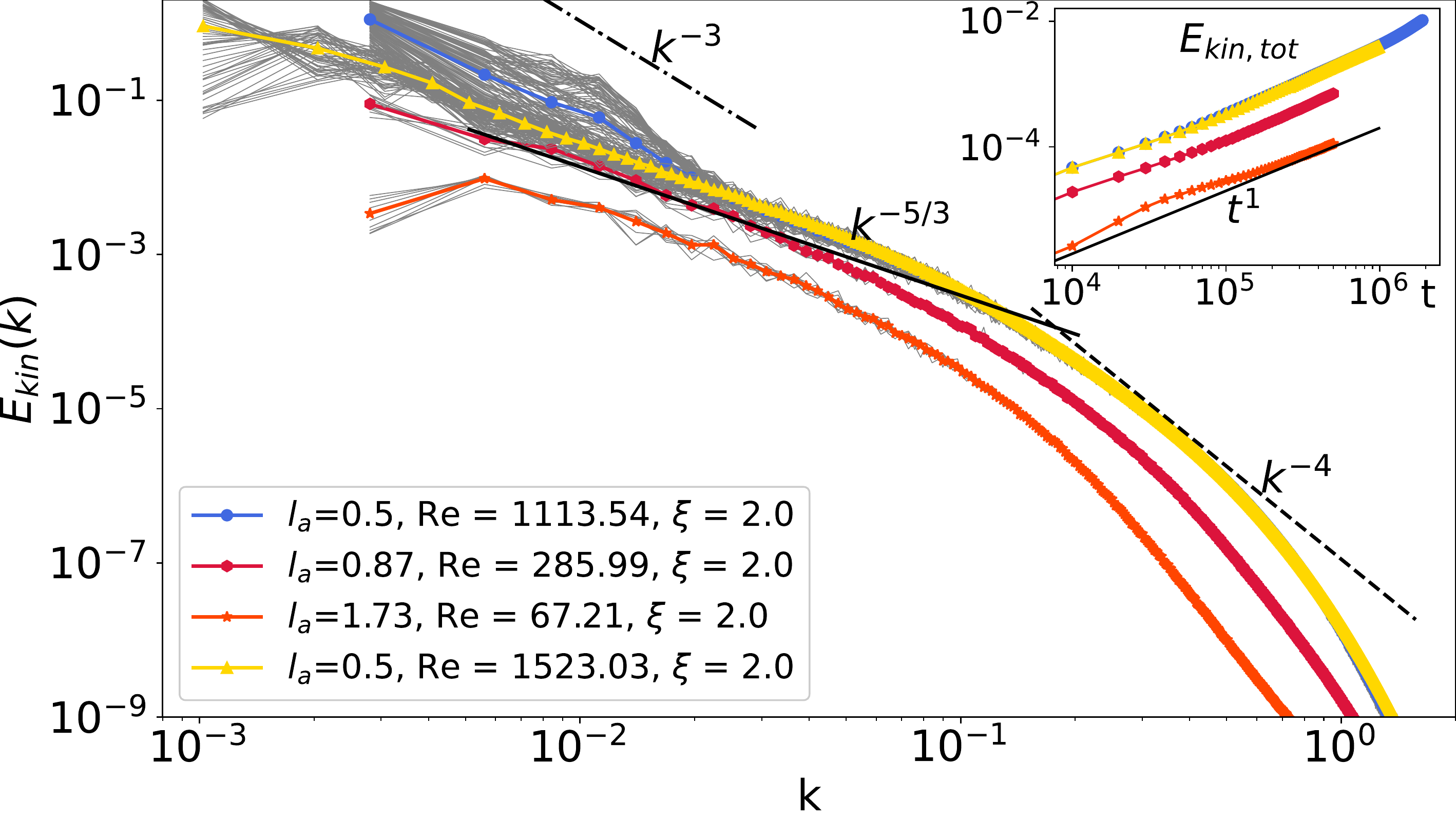}}
\put(-230,132){(b)}\\

\caption{\label{spectra} (a) Isotropic kinetic energy spectra rescaled by the kinetic energy at the active wavenumber $k_a$ identified by the intersection of the spectra with the dotted line. Black lines identify the $k^{-4}$ (solid) and $k^{-1}$ (dashed) scaling. The curves correspond to simulations of group E and D (Table~\ref{Runs}).  (b) Kinetic energy spectra and total kinetic energy density vs time (inset) in simulation units for systems of groups E and F characterized by the coexistence of active and inertial turbulence. Note the large Re numbers and $\xi$ and the different large scale scaling compared to the runs of panel (a) where inertial effects are negligible. See SM for the definitions of the isotropic kinetic, $E_{kin}(k)$, and elastic, $E_{el}(k)$, energy spectra} 
\end{figure*}
\begin{figure*} \centering
\resizebox{5.8cm}{!}{\includegraphics{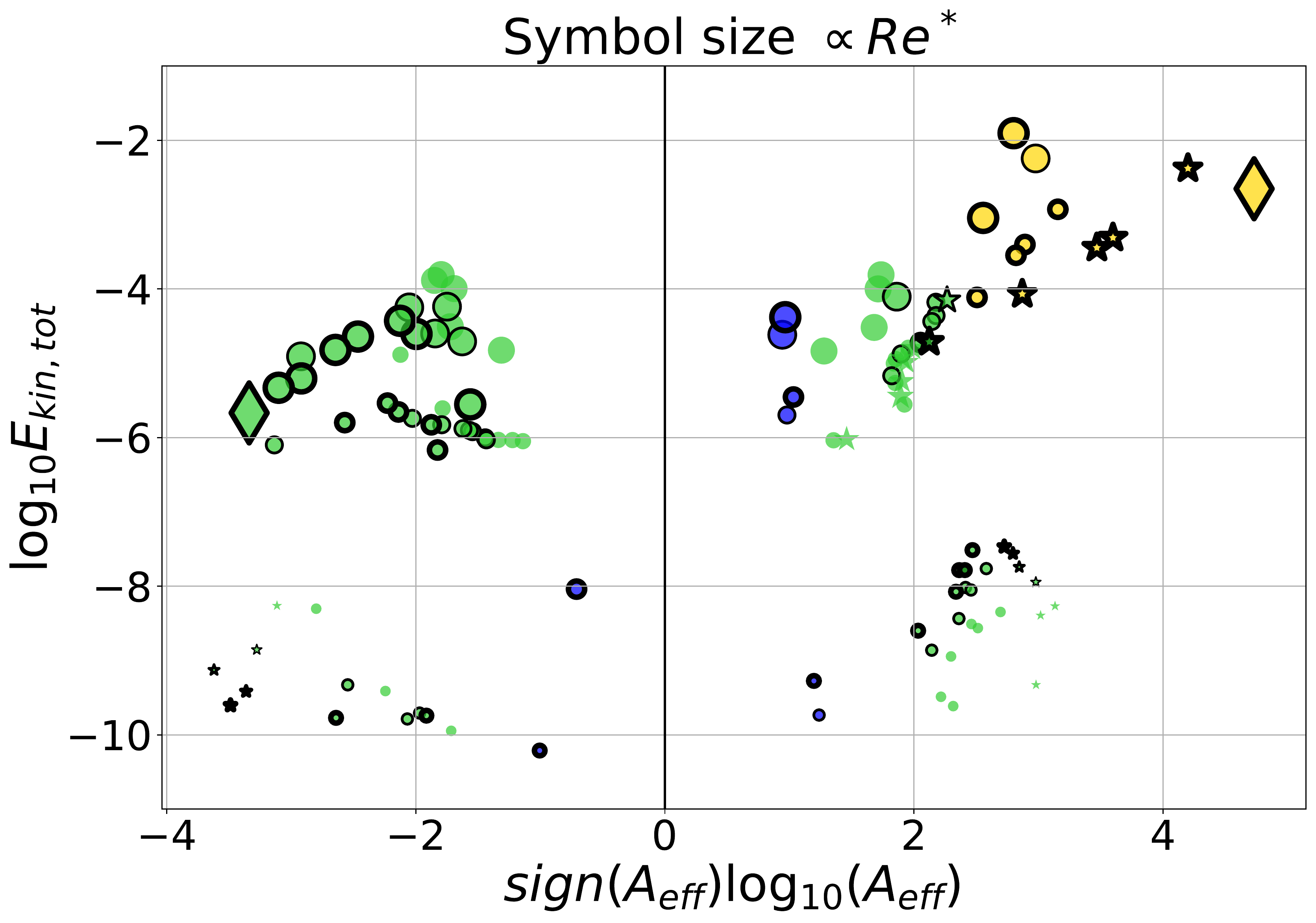}}
\resizebox{5.8cm}{!}{\includegraphics{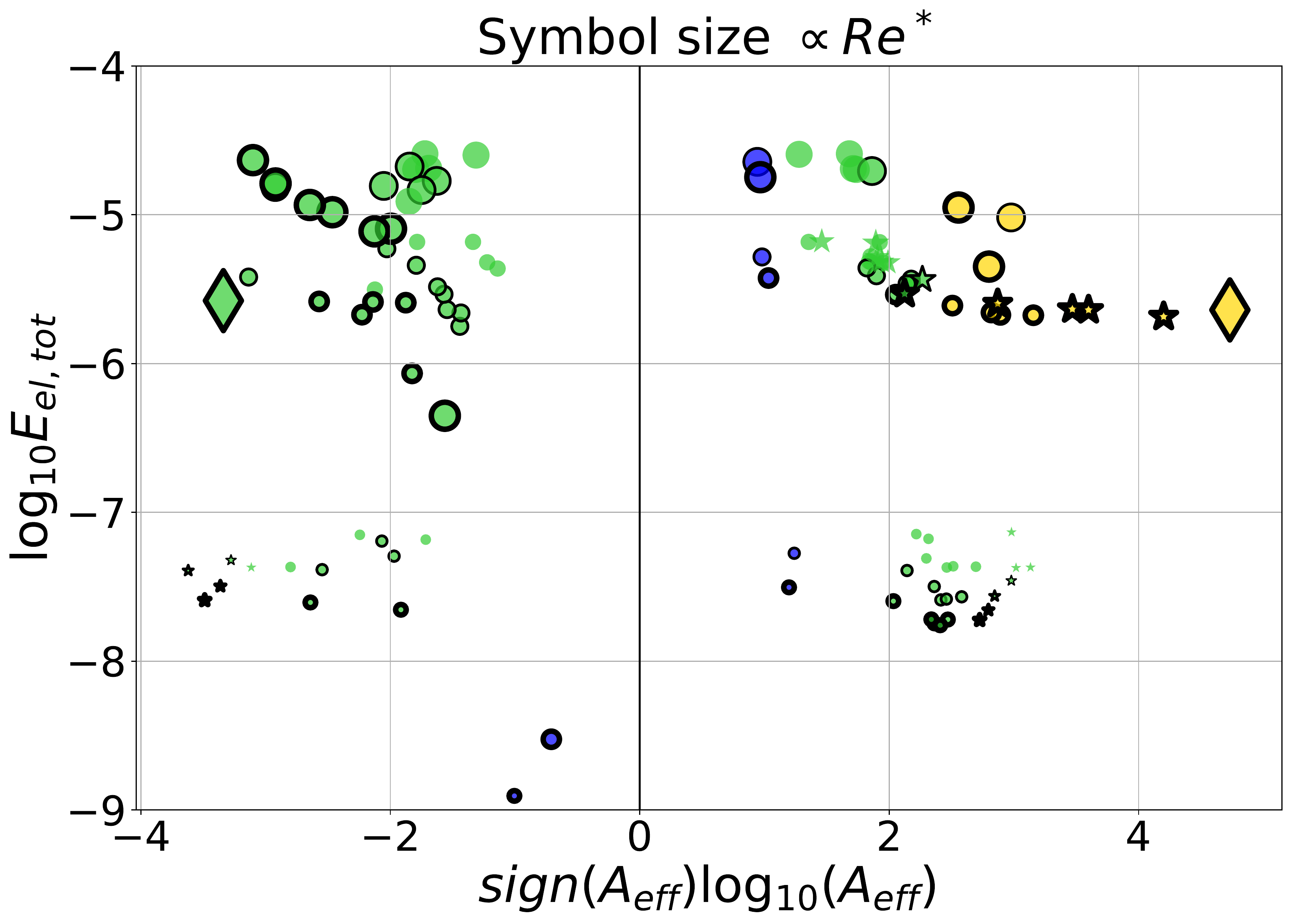}}
\resizebox{5.8cm}{!}{\includegraphics{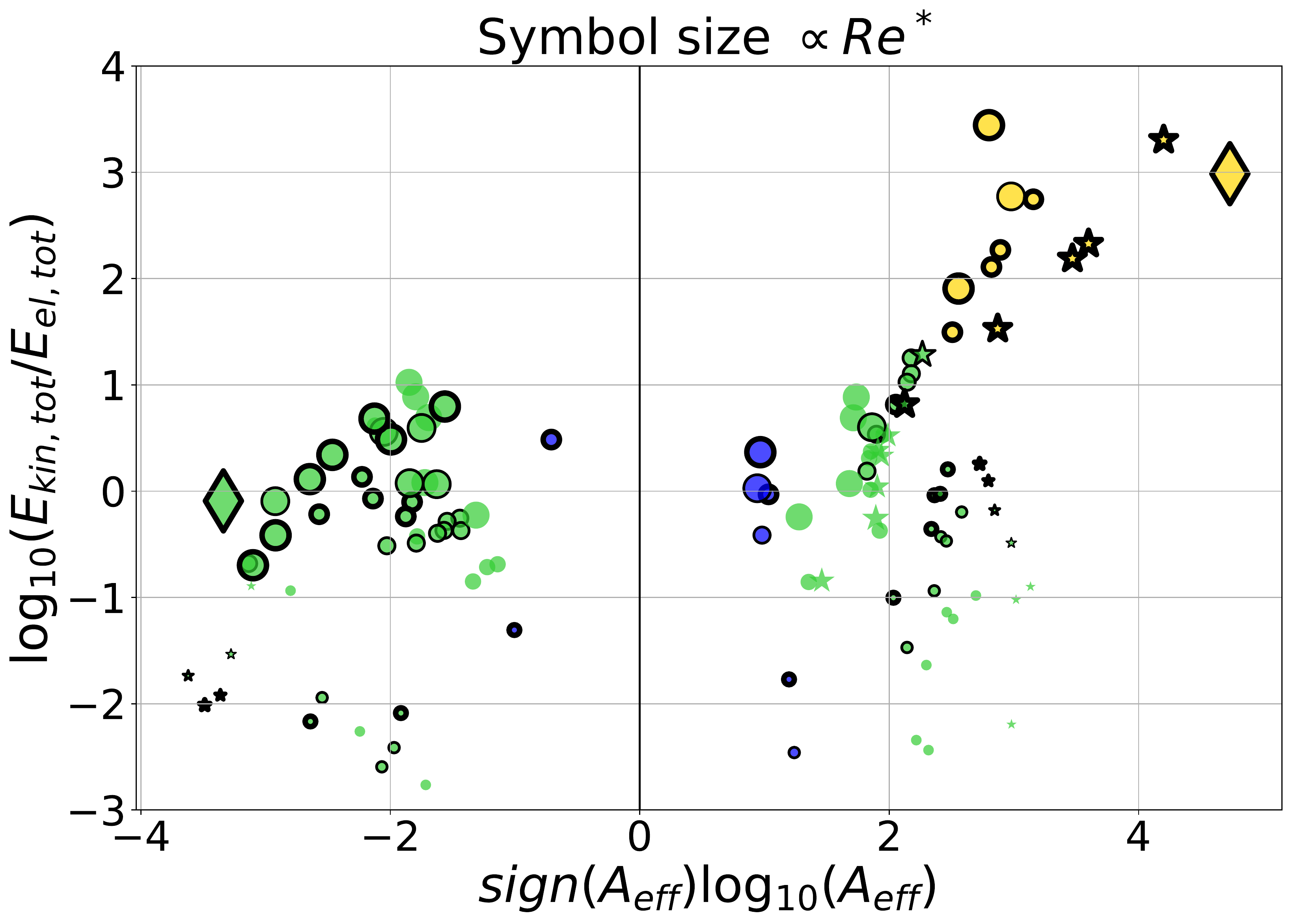}}
\put(-480,115){(a)}
\put(-320,115){(b)}
\put(-150,115){(c)}
\caption{\label{phase_diagram}  Logarithm of the total kinetic (a) and elastic (b) energy vs $A_{eff}$, that compares the characteristic size of the largest flow structures $\widetilde{L}_I$ with the effective active length scale $\widetilde{l_a}$ at which energy is injected into the system, for the simulations of Table \ref{Runs}. The different colors refer to the different type of steady states: non turbulent (blue), active turbulent (green), coupled active and inertial turbulence (orange). The size of the marker is proportional to the nominal Re number, $Re^*$, while the different symbols correspond to different resolutions $L=560$ (circle), $L=2240$ (star), $L=6144$ (diamond). The linewidth of the black symbols edge is proportional to $\xi$: $\xi=0$ (no  edge), $\xi=1$ (thin  edge) $\xi=2$ (thick  edge).  (c)  Ratio between the kinetic and elastic energy. Runs where active and hydrodynamic turbulence coexist are present only in the positive semiplane.} 
\end{figure*}
\begin{figure*} \centering
\resizebox{8cm}{!}{\includegraphics{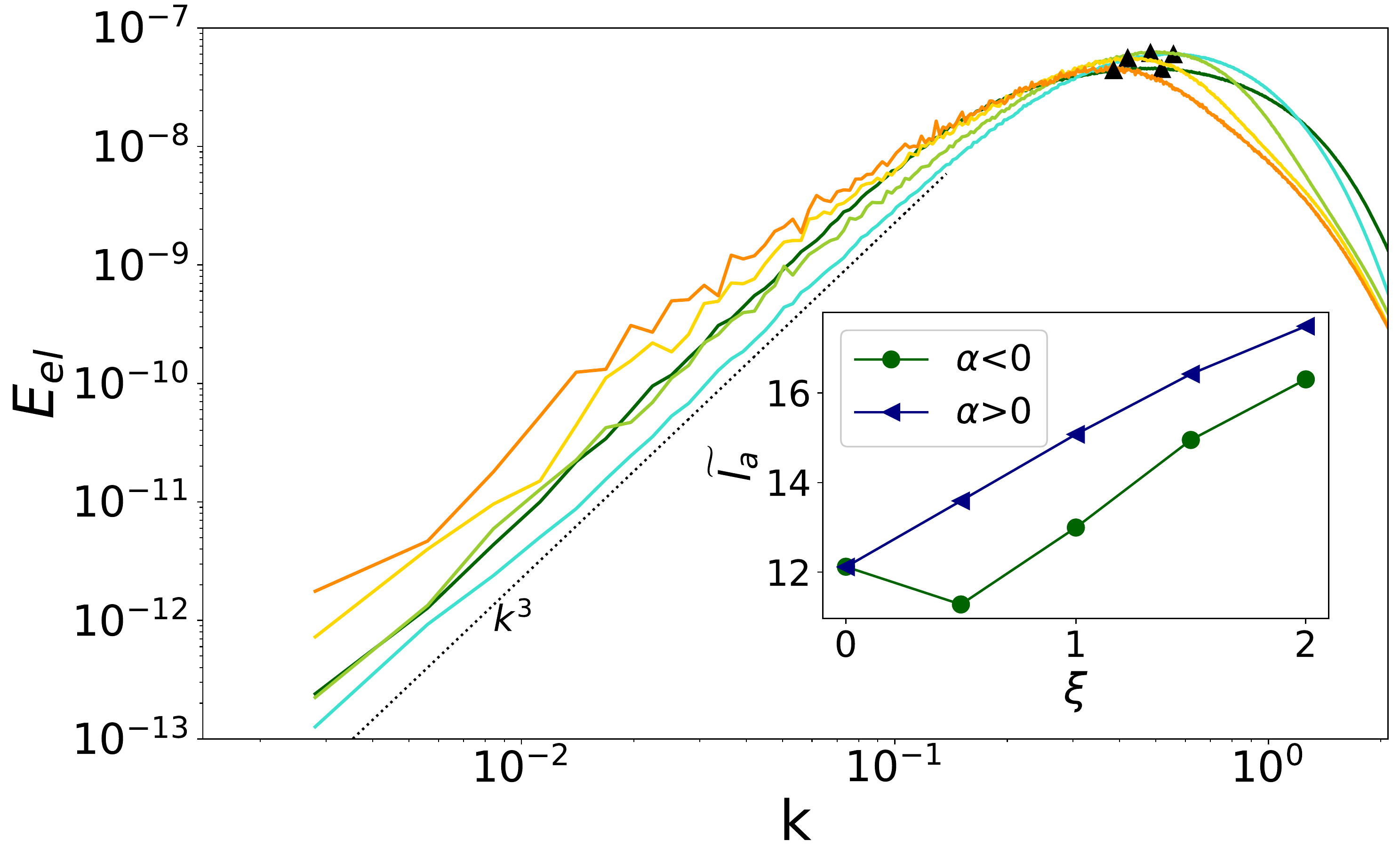}}
\resizebox{8cm}{!}{\includegraphics{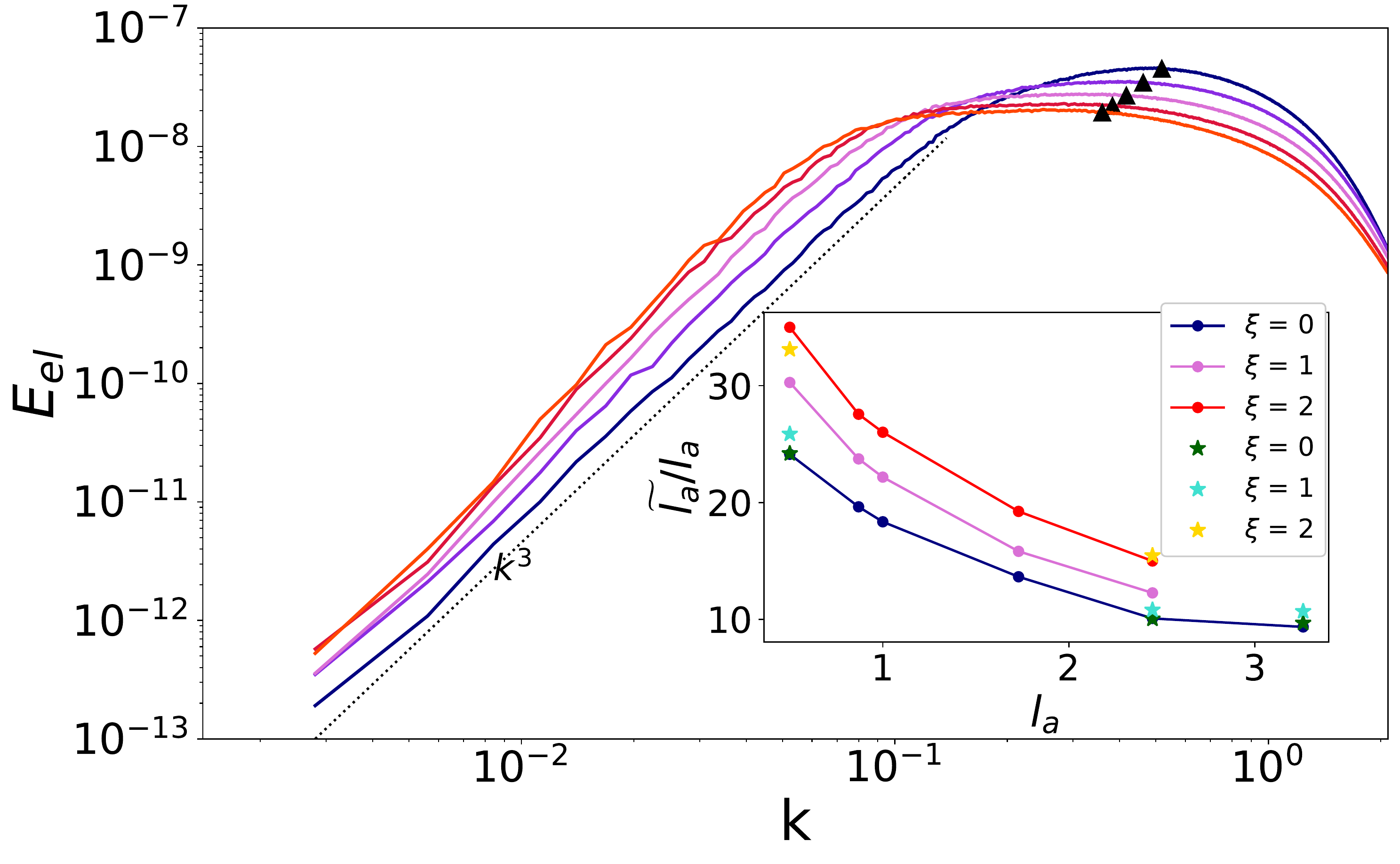}}
\put(-460,135){(a)}
\put(-230,135){(c)}\\
\resizebox{8cm}{!}{\includegraphics{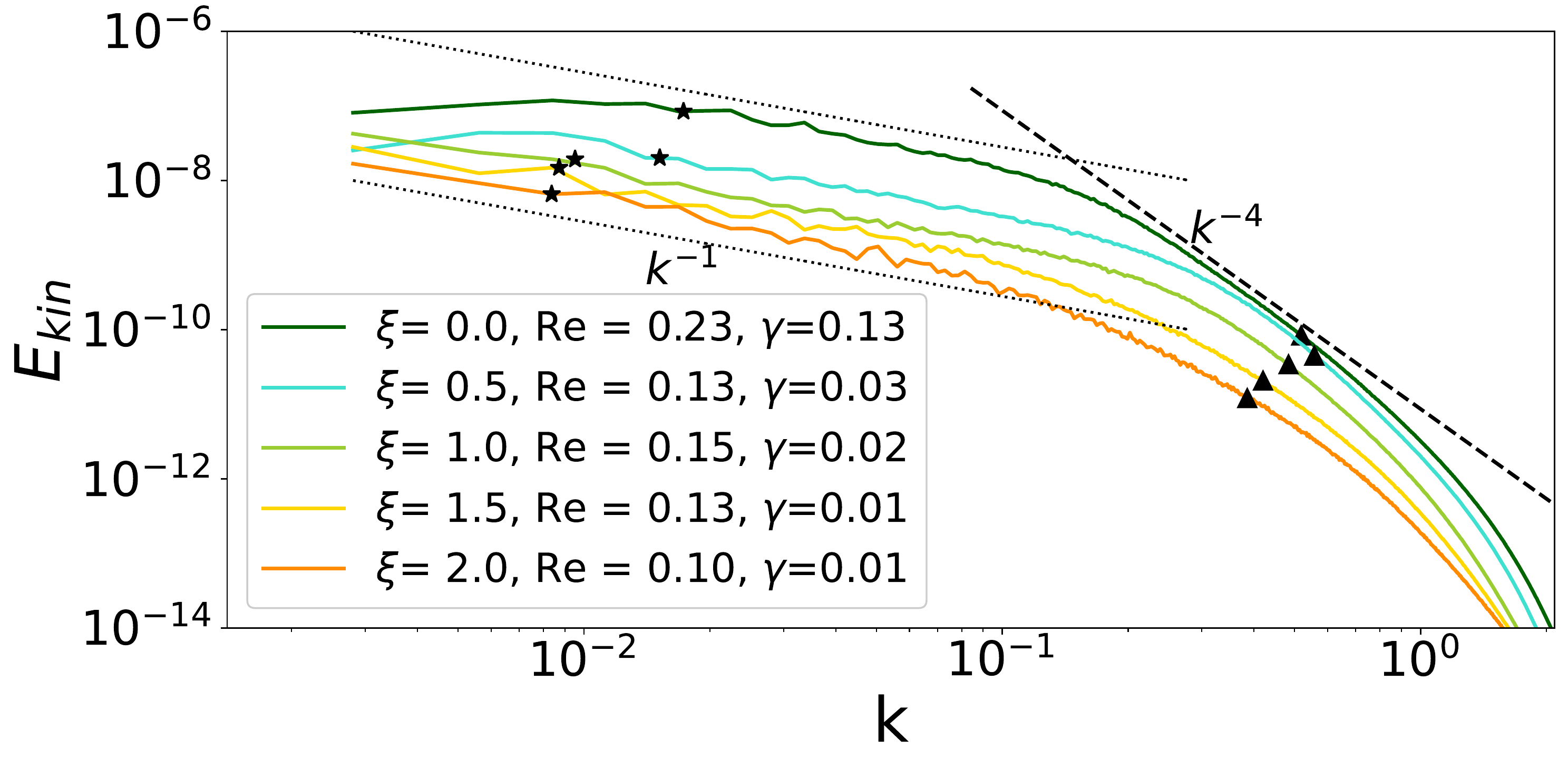}}
\resizebox{8cm}{!}{\includegraphics{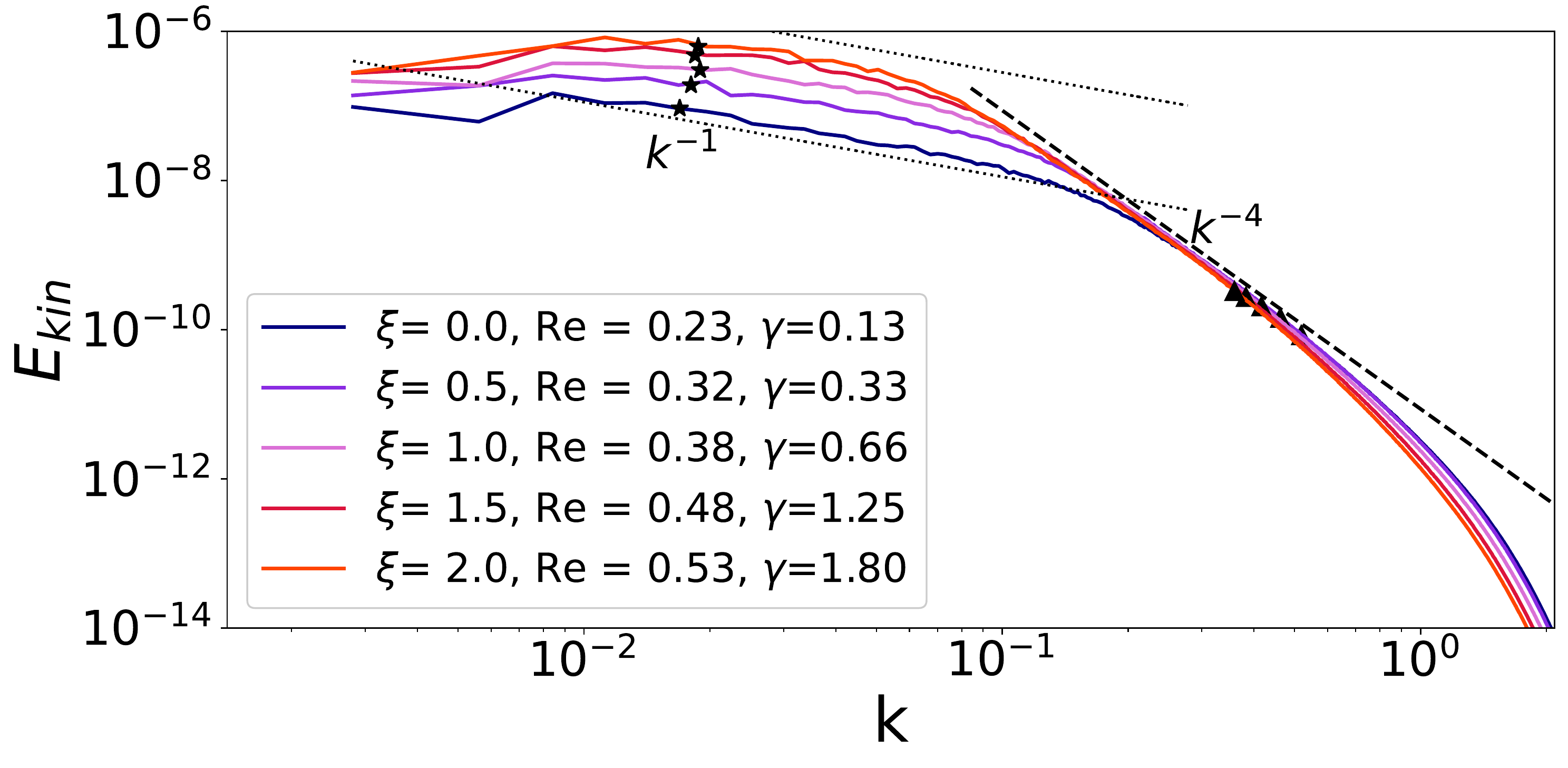}}
\put(-460,110){(b)}
\put(-230,110){(d)}\\
\caption{\label{explanation} Elastic (a),(c) and kinetic (b),(d) energy spectra in simulation units for  simulations of group D with  $l_a=-0.5$ (a)-(b) and $l_a=0.5$ (c)-(d) and different values of the flow-aligning parameter $\xi$ as reported in the legend of panel (b) and (d). The same color code is used in panel (a)-(b) and (c)-(d). The effective active wavenumber $k_a=2\pi/\tilde{l_a}$ is marked by black triangles while the wavenumber corresponding to the integral length scale, $\widetilde{L}_I$, is marked by a star in panel (b) and (d). The inset of panel (a) shows how the effective active length, $\tilde{l_a}$, varies with $\xi$ for  contractile and extensile nematics whose spectra are shown in this image. The inset of panel (c) shows how $\tilde{l_a}$ scaled by $l_a$ varies with $l_a$ for group A excluding those in a NT state. The increase of $\xi$ affects the spectra of a contractile and extensile nematics differently.} 
\end{figure*}
We integrate Eqs.~(\ref{QijEq})-(\ref{NSEq}) in a $2D$ periodic domain using a hybrid lattice-Boltzmann finite-difference method~\cite{Vincenzi15, Rorai21}. We impose a zero velocity field and a director field oriented along $x$ and randomly perturbed by a 10\% white noise as initial conditions. We explore a wide portion of parameter space (Table \ref{Runs}) and find that systems mostly evolve toward an active turbulent state, Fig.~\ref{fields}(a), or a state where active and hydrodynamic turbulence coexist, Fig.~\ref{fields}(b)-(c). 
In both cases the multiscale nature of the flow is revealed by the presence of structures of different sizes, however, only when inertial turbulence is triggered the largest structures reach the system size.

We classify as {\it active turbulent}, systems  that reach a statistical steady state characterized by energy spectra with the scaling predicted in Refs.~\cite{Giomi15, Joanny20}.
In these systems, the contribution of inertia is marginal, see the low Re numbers  in Fig.~\ref{spectra} (a).  Fig.~\ref{spectra}(a) displays  the   kinetic energy spectra, $E_{kin}(k)$, for active turbulent flows:  the predicted $k^{-4}$ scaling is always visible unlike the $k^{-1}$ large-scale scaling, observed only for sufficiently low Re numbers and high resolutions~\footnote{Previous studies~\cite{Urzay17} that included the inertial term in Eq. (\ref{NSEq}) also found $E_{kin}(k)$ to differ from the $k^{-1}$ power-law at large scales}, see SM for further details on the scaling. 
As opposed to the case in which $Re=0$ Ref.~\cite{Joanny20}, the peak of of the elastic energy  spectrum does not coincide  with the one of the kinetic energy spectrum: see in Fig. \ref{spectra}(a) the position of the effective active wave number ${k}_a=2\pi/\tilde{l}_a$, corresponding to the first moment of the elastic energy spectrum, $E_{el}(k)$. 

Inertial and active turbulence coexist when the flows develop  a condensate state composed  by  a system spanning vortex pair~\cite{Chertkov07, Laurie14}, Fig. \ref{fields}(c). Condensation is a finite-size effect expected in forced $2D$ hydrodynamic turbulence  when large scale drag is absent~\cite{Alexakis18}. The convergence to the condensate state is signaled by ({\it i}) the formation of large scale structures, Figs.~\ref{fields}(b)-(c), ({\it ii}) the scaling of the total kinetic energy density, $E_{kin, tot} = \int_0^{\infty} E_{kin}(k)dk$, as $t^1$~\cite{Chertkov07}, see Fig.~\ref{spectra}(b) inset, and ({\it iii}) a large-scale scaling of $E_{kin}(k)$ consistent with $k^{-5/3}$  and $k^{-3}$ at the largest wavenumbers where the finite size effects are felt \cite{Chertkov07}, see Fig.~\ref{spectra}(b). 
Because of the steady growth of the largest modes these systems have not reached a steady state. However, the small and intermediate scales have converged: compare in Fig.~\ref{spectra}(b) the spectra averaged over the last time steps (in color) with the instantaneous spectra in the time window where the average is performed (gray). The signature of active turbulence is recognizable at smaller scales where the kinetic energy scales as $k^{-4}$, Fig.~\ref{spectra}(b).

We explore parameter space  systematically, beyond  Figs. \ref{fields} and \ref{spectra}, covering flow-tumbling $\xi<1$ and flow-aligning $\xi>1$ regimes for both $\alpha<0$ and  $\alpha>0$ and different active lengths, system sizes, and Re numbers: see Table~\ref{Runs}. Active nematics reach three different states: an active turbulent (AT) state  as in Fig.~\ref{fields}(a) and \ref{spectra} (a), coupled active and hydrodynamic turbulence (AHT state) as in Fig.~\ref{fields}(b)-(c) and \ref{spectra}(b), and a non-turbulent (NT) state.
We classify as {\it non turbulent}, systems that do not display a Gaussian velocity distribution and a well developed isotropic kinetic energy spectrum. These correspond to  slowly-varying time-dependent flows whose topological features are walls that reconfigure in time~\cite{Giomi11, Giomi14}, see SM. 

To identify boundaries in phase-space between the NT and the AT states and between the AT and the AHT states,  Fig.~\ref{phase_diagram} displays  the total kinetic energy density and the total elastic energy density, $E_{el, tot} = \int_0^{\infty} E_{el}(k)dk$, for all the runs of Table \ref{Runs} as a function of an effective active parameter $A_{eff}=2q_0 \widetilde{L}_I^2/(\widetilde{l_a}^2\Gamma\eta)$ where $\widetilde{L}_I$ is the first moment of the kinetic energy spectrum and represents the characteristic size of the largest flow structures, while $\widetilde{l_a}$ represents the scale at which energy is injected through active forcing; see  SM for further considerations. In Fig. \ref{phase_diagram} the data are distinguished by the state of the system (different colors), the value of $Re^*$ (different symbol sizes), by the system size (different symbols) and by  the flow-aligning parameter (different line-width); see SM for a visualization of the data grouped by $\xi$. The plot shows that the three dynamical states  can be separated by equal $A_{eff}$ lines.  
Fig.~\ref{phase_diagram} reveals striking asymmetries between the negative and positive semiplane: ({\it i}) no AHT runs are found in the negative semiplane, ({\it ii}) the transition from the NT to the AT state occurs for smaller $A_{eff}$ for contractile nematics and for different type of flows, see SM, ({\it iii}) the largest  kinetic energies are associated to the lowest  $\xi$ for a contractile nematic, the opposite is true for an extensile nematic and the runs that display an inverse energy cascade have $\xi\ge1$, ({\it iv}) $E_{kin,tot}$  increases with  $A_{eff}$, except for the most energetic flows in the negative semiplane where it decreases with $|A_{eff}|$,  ({\it v}) $E_{el,tot}$, despite showing a reduced variability compared to $E_{kin,tot}$, tends to decrease with $|A_{eff}|$ except for a contractile nematics with $\xi =2 $. Fig.~\ref{phase_diagram} (c) displays the ratio between the kinetic and elastic energy in logarithmic scale, showing that generically for contractile active flows the elastic energy dominates over the kinetic energy, while the opposite is true for extensile active flows.

The kinetic and elastic energy spectra for extensile and contractile nematics, Fig. \ref{explanation},  show that 
flows characterized by a positive and negative activity respond differently to an increase of the flow-aligning parameter. An increase of $\xi$ results in an increase of the effective active length $\tilde{l_a}$, that is, an increase in the correlation length of the $Q_{ij}$ tensor~\cite{Thampi15}. This  is observed  both for contractile and extensile nematics and reveals that the characteristic length of the nematic structures increases with $\xi$, insets of Fig.~\ref{explanation} and SM.
However, $E_{el}(k)$ shows that $\xi$ affects the redistribution of energy across scales differently for a contractile and extensile nematics. An increase of $\xi$ causes  the peak of $E_{el}(k)$ to sharpen for a contractile nematics, Fig. \ref{explanation} (a), while in an extensile nematics, $E_{el}(k)$ flattens considerably indicating that the highly energetic structures span a wider range of scales, Fig. \ref{explanation} (c); this is reflected in a wider range of wavenumbers where the kinetic energy scaling follows the $k^{-4}$ power-law. The broadening of the $k^{-4}$ power-law to larger scales results in an increase of the kinetic energy as also evidenced by an increase of the ratio $\gamma$ between the kinetic and elastic energy (legend of Fig. \ref{explanation}(d)). This mechanism, intrinsic to the coupled effect of an extensile nematics and a large  $\xi$, can trigger inertial turbulence through the inverse kinetic energy cascade if the Re number of a reference $\xi=0$ flow is large enough.

To conclude, we have unveiled a mechanism to rationalize the coexistence of active and inertial turbulence for  active extensile and flow-aligning nematics. Remarkably, our finding that large Re number flows develop for  extensile and flow-aligning nematics mirrors  the experimental results of superfluid-like behavior for  elongated pusher-like bacteria \cite{Lopez15}. However, drawing closer analogies  is challenging due to the lack of  data  on the flow characteristics, {\it e.g.} large scale coherent motion vs active turbulent flow, when a negligible viscosity is measured inside the rheometer.

\begin{acknowledgments}
We acknowledge funding from the European Union's Horizon 2020 research and innovation program under the Marie Sklodowska-Curie grant agreement N 754462. I.P. acknowledges support from Ministerio de Ciencia, Innovaci\'on y Universidades (Grant No. PGC2018-098373-B-100/FEDER-EU), DURSI (Grant No. 2017 SGR 884), and SNSF (Project No. 200021-175719). This work was supported by a grant from the Swiss National Supercomputing Centre (CSCS) under project ID s1079.
\end{acknowledgments}
%

\end{document}